\documentstyle[prl,aps]{revtex}
\input epsf
\begin{document}
\draft
\twocolumn[\hsize\textwidth\columnwidth\hsize\csname @twocolumnfalse\endcsname

\title{Metal-Insulator Transition in One-dimensional Hubbard Superlattices}

\author{Thereza Paiva and Raimundo R.\ dos Santos}

\address{Instituto de F\'\i sica, 
         Universidade Federal Fluminense\\
         Av.\ Litor\^anea s/n,
         24210-340 Niter\'oi RJ, Brazil}
\date{\today}
\maketitle

\begin{abstract}
We study the Metal-Insulator transition in one-dimensional 
Hubbard superlattices (SL's), modelled by a repeated pattern 
of repulsive (i.e., positive on-site coupling) and free sites.
The evolution of the local moment and of the charge gap (calculated
from Lanczos diagonalization of chains up to 18 sites), together
with a strong coupling analysis, show that the electron density
at which the system is insulating increases with the size of the 
free layer, relative to the repulsive one.
In the insulating state, the mechanism of interaction between fermions 
separated by a free layer is the analog of superexchange, 
and the charge gap display universal features.
\end{abstract}
\vskip2pc]

After many decades, the Metal-Insulator transition (MIT) still poses a
great challenge, from both theoretical and experimental points of view\cite{Mott90}.
For instance, one of the fundamental difficulties with MIT's is the
identification of an order parameter which, in analogy with other 
(e.g., magnetic) phase transitions, would lead to {\em spatial-dependent}
correlation functions and correlation lengths\cite{Talstra}. 
Nevertheless, considerable progress has been achieved in understanding 
and describing this phenomenon.
MIT's can be driven by electronic correlations (the Mott transition), 
by disorder (Anderson transition), or by structural instabilities. 
The Mott transition, which is the one we discuss here, can take place
as one varies any of the thermodynamic parameters, such as electron 
density, temperature, or the strength of correlation energy.

In the quest for a deeper understanding, it is important to gather 
as much information as possible about systems undergoing a MIT, 
by means of simplified models amenable to well-controlled 
approximations, and the Hubbard model provides a nice testing ground
for new ideas. 
In effect, the ground state of the half-filled homogeneous chain 
corresponds to an insulating  Spin-Density--Wave (SDW) state, which 
resembles the N\'eel state, but with power-law decay of 
correlations\cite{LiebWu};
its strong-coupling limit describes localized spins coupled
antiferromagnetically through a Heisenberg exchange interaction.
Away from half-filling the ground state is metallic,
with weak SDW correlations\cite{LiebWu,Mucio94}.
In two dimensions, the concurrent van Hove 
singularity and nesting of the Fermi surface at half-filling give rise to 
an insulating antiferromagnetic (AFM) ground state for any finite on-site 
Coulomb repulsion $U$. 
As one dopes away from half-filling, the system undergoes a transition 
to a paramagnetic metal\cite{Hirsch85,Hirsch89}, with weak short-ranged 
incommensurate spin-spin correlations\cite{Moreo90,spiral}.
Modifying the band structure through the addition of a second-neighbor
hopping term does not seem to displace the transition point
from half-filling\cite{Duffy95}.
Considerable insight into the MIT problem should therefore be gained 
by seeking instances in which such displacement can occur.

Motivated by the oscillatory exchange coupling in magnetic metallic 
multilayers\cite{MMM}, we have recently studied the effects 
of electronic correlations in superlattices (SL's), through
a one-dimensional Hubbard-like model.
In spite of its simplicity, a number of remarkable features were
found, in marked contrast with the otherwise homogeneous system\cite{tclp1}:
Local moment weight could be transferred from repulsive to free sites; 
SDW quasi-order was wiped out as a result of frustration; 
and strong SDW correlations (in a subset of sites) could set in above 
half-filling.
Since the combination of strong correlations with SL structures can
modify the magnetic properties in such non-trivial way, transport 
properties should, most likely, be also affected.
With this in mind, here we investigate the MIT transition in this 
one-dimensional Hubbard superlattice model.

The model consists of a periodic arrangement of $L_U$ sites 
(``layers'') in which the on-site coupling is repulsive,
followed by $L_0$ free ({\it i.e.,} $U=0$) sites, such that the Hamiltonian 
is written as
\begin{equation}
\label{Ham}
{\cal H}=-t\sum_{i,\ \sigma}
\left(c_{i\sigma}^{^{\dagger}} c_{i+1\sigma}+\text{H.c.}\right) 
+ \sum_i U_i\ n_{i\uparrow}n_{i\downarrow}
\end{equation}
where, in standard notation, $i$ runs over the sites of a one-dimensional 
lattice, $c_{i\sigma}^{^\dagger}$ ($c_{i\sigma}$)
creates (annihilates) a fermion at site $i$ in the spin state 
$\sigma=\uparrow\ {\rm or}\ \downarrow$, and
$n_i=n_{i\uparrow}+n_{i\downarrow}$, with 
$n_{i\sigma}=c_{i\sigma}^{^\dagger}c_{i\sigma}$.
The on-site Coulomb repulsion is taken to be 
site-dependent:
$U_i=U>0$, for sites within the repulsive layers, and $U_i=0$
otherwise; throughout this paper, numerical values of $U$ will be given
in units of $t$.  
It is important to notice that the SL structure breaks particle-hole 
symmetry even on bipartite lattices, as it can be seen through a simple 
Hartree-Fock argument:  
The exchange splitting is zero on free sites and non-zero on the 
repulsive ones, thus giving rise to a non-uniform shift in the 
symmetric one-body local density of states. 
Differently from the homogeneous system, particle-hole symmetry cannot 
be restored by a uniform shift in the chemical potential.

The MIT in this model is studied numerically in the canonical ensemble.
In addition to the numbers of sites, $N_s$, and electrons, $N_e$, another important parameter is the number of periodic cells, $N_c=N_s/N_b$, for a basis with $N_b=L_U+L_0$ sites; 
in some cases, we were able to reach lattices with $N_s=18$ sites.
The ground state $|\psi_0\rangle$ and energy of Eq.\ (\ref{Ham}) 
are obtained with the aid of the Lanczos algorithm\cite{Roomany80,Gagliano86,Dagotto94}: 
Starting with a trial state, the Hamiltonian is used to generate a second
state, orthogonal to the first, so one ends up with a $2\times 2$ 
representation for the Hamiltonian\cite{Gagliano86}.
The diagonalization is trivial, leading to an estimate for the ground state 
and energy, which are used as new inputs for the subsequent iteration.
This process is repeated until numerical convergence for the ground state 
energy has been achieved. 
In order to improve extrapolations to the thermodynamic limit, boundary 
conditions were chosen such that the Fermi momentum was always one of the 
allowed $k$-values\cite{Jullien82}.
The influence on the results can be assessed by simultaneously fixing 
system size, occupation and SL configuration, and performing calculations 
for different boundary conditions:
the ground state energy turned out to be quite insensitive to the
condition imposed (at most of the order of 1\%) and similarly for the 
local moment.
When dealing with the charge gap (see below), the above mentioned choice 
has the additional advantage of providing a trend (with system size) free 
from oscillations\cite{Jullien82}.

We assess whether the system is metallic or insulator by calculating the local moment and the charge gap, as we now discuss in turn.
The local moment at site $i$ is defined as 
\begin{equation}
\label{moment}
\langle S_i^2\rangle={3\over 4} \langle m_i^2\rangle,
\end{equation}
where 
$m_i\equiv n_{i\uparrow}-n_{i\downarrow}$; 
ensemble averages should be understood as ground state averages, 
$\langle\ldots\rangle\equiv\langle\psi_0|\ldots|\psi_0\rangle$.
Being a measure of both magnetism and degree of itinerancy of 
the system, the local moment is useful in the investigation of the 
metallic or insulating character of the ground state.
Indeed, in the case of a homogeneous lattice, for a finite value of the 
on-site repulsion $U$, the local moment is sharply peaked 
at half-filling; 
in the completely localized limit, $U=\infty$, $\langle S^2 \rangle=3/4$.
Above half filling, $\langle S_i^2\rangle$ decreases due to an increase 
in the double occupancy of the sites.
Unlike the magnetization (or even the sublattice magnetization), 
$\langle m_i\rangle$, which vanishes identically
on a finite system due to the lack of spontaneous symmetry breaking,
the local moment is always non-zero on a homogeneous lattice, except
for $\rho=0$ and 2. 

\begin{figure}
\epsfxsize=7,5cm
\begin{center}
\leavevmode
\epsffile{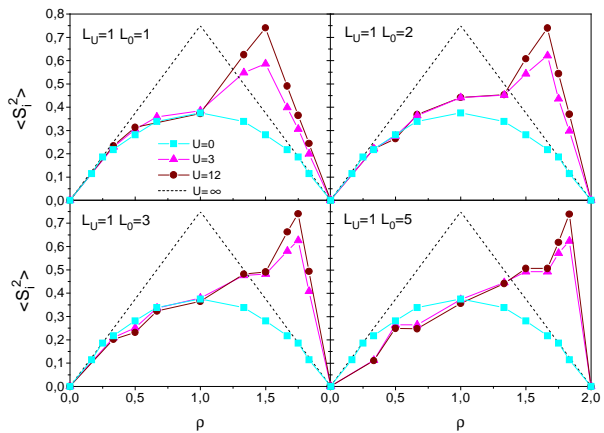}
\caption{Local moment at a repulsive site versus $\rho$ for 12 site
chains with $L_U=1$ and $L_0= 1$ (a), 2 (b), 3 (c), and 5 (d).
Squares and dotted lines respectively represent data for $U=0$ and $\infty$
for the homogeneous system; triangles and circles respectively represent 
data for superlattices with $U=3$ and 12.} 
\label{lu1xro} 
\end{center}
\end{figure}

For a SL the site-dependent Coulomb repulsion leads to a non-uniform
distribution of local moments throughout the lattice\cite{tclp1}.
Figure\ \ref{lu1xro} shows typical plots of the density dependence of the 
local moment at a repulsive site, for lattices with $N_s=12$.
As the SL configuration is changed, the maximum value of $\langle S_i^2\rangle$
still approaches 3/4 as $U$ is increased; 
its position is displaced to higher fillings, without showing any
$U$-dependence. 
By analogy with the homogeneous case, one is led to identify this peak 
position with the density, $\rho_I$, at which the system becomes an insulator;
Fig.\ \ref{lu1xro} then shows that $\rho_I$ increases continuously 
with $L_0$, for fixed $L_U$. 
To see how this comes about, we should ask ourselves how can the SL be filled 
up with electrons in a way to obtain maximum hopping hindering and largest
moment on the repulsive sites. 
In strong coupling this is equivalent to pin the electrons, and is achieved 
by placing two on each of the free sites (thus rendering them magnetically
inert) and one on each of the repulsive sites (maximum polarization);
this leads to
\begin{equation}
\rho_I ={{2L_0+L_U} \over {L_0+L_U}}.
\label{rhoi}
\end{equation}
The maxima position of Fig.\ \ref{lu1xro} are given exactly by Eq.\ (\ref{rhoi}),
with $L_U=1$, and $L_0=1,2,3,$ and 5. 
Configurations with $L_U>1$ follow the same pattern, and reflect the
fact that the layer thicknesses come into the definition of $\rho_I$ 
only through the ratio $L_U/L_0$.   
Also, note that in the limit of a uniform lattice, $L_0\to 0$, one has
$\rho_I\to 1$, thus recovering the insulating behavior exactly at half filling.

The above discussion allows us to 
perform a strong coupling perturbation theory analysis\cite{Emery76}.
Let us start with the dimerized SL configuration; the relevant Hilbert 
subspace is the one with repulsive and free sites being respectively 
singly and doubly occupied. 
The first non-trivial contribution from the hopping term 
comes in fourth order,
and gives rise to an effective Heisenberg Hamiltonian
\begin{equation}
\label{t4}
{\cal H}_{\rm eff}=J_2 \sum_{\left[k,\ell\right]} {\bf S}_k\cdot {\bf S}_\ell,
\end{equation}
where $J_2=8t^4/|U|^3$, and $k$, $\ell$ are repulsive sites (i.e., {\em next-nearest} neighbors on the original dimerized lattice); trivial 
constants have been dropped from Eq.\ (\ref{t4}).
Thus, the (magnetically inert) free sites intermediate the Heisenberg 
exchange interaction between moments on the repulsive sites; the analogy
with the usual superexchange mechanism involving non-magnetic atoms
arises here quite naturally.
These arguments can be straightforwardly generalized to other
configurations, with suitable changes: 
(i) within a repulsive layer ($L_U>1$), spins interact via the 
usual second order Heisenberg exchange coupling $J_1=4t^2/|U|$;
(ii) the first non-trivial coupling across a free layer of thickness
$L_0$ only appears in $2(L_0+1)$-th order, and the exchange coupling 
becomes $\sim t^{2(L_0+1)}/|U|^{2L_0+1}$. 
The ensuing Heisenberg model is again defined only on the repulsive sites, with non-uniform, but periodic, exchange couplings. Away from strong coupling, numerical data are consistent with this picture\cite{tclp1}.

Further evidence in favor of the MIT being located at $\rho_I$ comes 
from the analysis of the charge gap, defined by
\begin{equation}
\Delta_c = E(N_c, N_e+1) + E(N_c,N_e-1)-2E(N_c,N_e),
\label{cgap}
\end{equation}
where, for a given SL configuration, $E(N_c,N_e)$ is the ground 
state energy for a chain with $N_c$ periodic cells and $N_e$ electrons.  
This gap is a measure of the charge excitation spectrum 
(i.e., the cost of adding extra particles):
When extrapolated to the thermodynamic limit, a non-zero value indicates
an insulating state, whereas a zero value corresponds to a metallic state. 
Before discussing our results, a comment is in order at this point. 
The Drude weight, which is proportional to the the second
derivative of the ground state energy with respect to the magnetic flux 
through the ring\cite{Kohn64,Shastry90,Scalapino92}, is an 
elegant test of the metallic or insulating character of the system:
It yields impressive results when Bethe-Ansatz data (corresponding to 
arbitrarily large lattices) are used\cite{Shastry90}.
For small systems, however, extrapolations with the charge gap appear to be
more conclusive than those with the Drude weight\cite{Penc94,Arrachea96}.
    
\begin{figure}
\epsfxsize=7,5cm
\begin{center}
\leavevmode
\epsffile{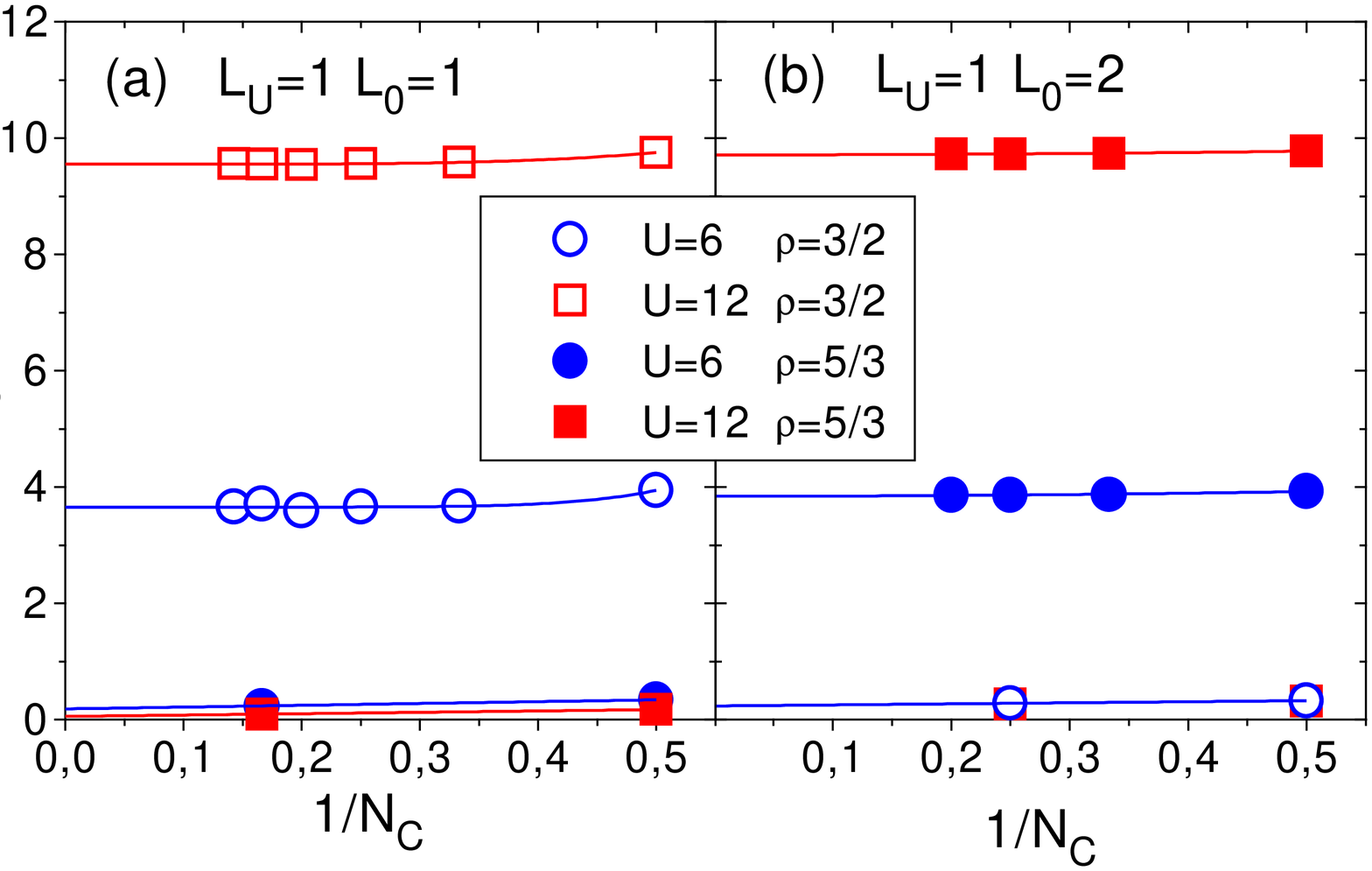}
\caption{Charge gap vs.\ the inverse number of cells for the SL configurations
$L_U=L_0=1$ (a) and $L_U=1,\ L_0=2$ (b), and densities $\rho=3/2$ 
(empty symbols) and $\rho=5/3$ (filled symbols). Circles and squares denote
$U=6$ and $U=12$, respectively. The lines are least-squares fits to the data.}
\label{cgap1} 
\end{center}
\end{figure}

Figure \ref{cgap1}(a) shows charge gap data for the dimerized configuration,
$L_U=L_0=1$.
In this case $\rho_I=3/2$, and extrapolation (through least-squares fits) 
of data towards $N_c\to\infty$ yields finite values for $\Delta_c$, for 
both $U=6$ and 12, confirming the insulating character of this filling factor.
By contrast, for $\rho=5/3$, the data extrapolate to $\Delta_c\simeq 0$,
for both values of $U$, signaling a metallic state;
for clarity, the only metallic case shown in Fig.\ \ref{cgap1}(a) corresponds 
to $\rho=5/3$, but this is a representative example for all $\rho\neq\rho_I$.
For the configuration $L_U=1,\ L_0=2$, $\rho_I$ is now 5/3, and Fig.\ 
\ref{cgap1}(b) shows that the roles of fillings 5/3 and 3/2 have been
reversed with respect to that of Fig.\ \ref{cgap1}(a).

\begin{figure}
\epsfxsize=7,5cm
\begin{center}
\leavevmode
\epsffile{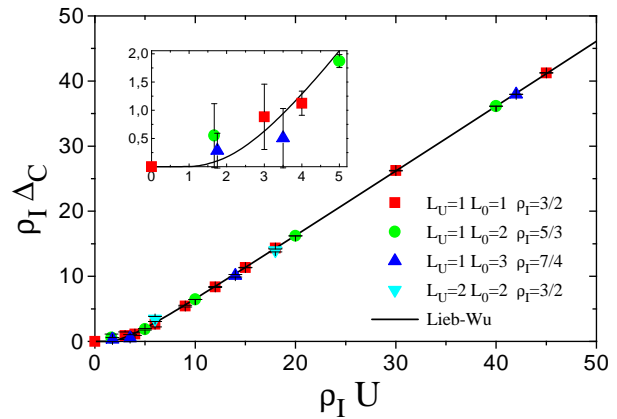}
\caption{Scaled extrapolated--charge-gap versus scaled on-site energy. 
Data for the configurations ($L_U,L_0$) are represented as follows:
(1,1) by squares; (1,2) by circles; (1,3) by up-triangles; and (2,2) by 
down-triangles. 
The full line is the Mott-Hubbard gap for the homogeneous system,
calculated from the Lieb-Wu solution.
The inset shows the intermediate coupling region.}
\label{liebwu}
\end{center}
\end{figure}

Similarly to the homogeneous system, the gap increases with 
$U$. 
This similarity, however, goes beyond a simple qualitative trend: 
If both the extrapolated charge gap {\em and} the on-site coupling $U$ are scaled 
by $\rho_I$, strong coupling data for different $\rho_I$ fall on a
universal function, as shown in Fig.\ \ref{liebwu}. 
This function, in turn, reduces to the Mott-Hubbard gap for the 
homogeneous system ($\rho_I=1$), known for any $U$ through the 
Bethe {\em ansatz} solution\cite{LiebWu,Kawakami89}; its asymptotic forms are  
$\Delta_c\sim (8\sqrt{tU}/\pi)\exp{\left(-2\pi t/U\right)}$, in weak
coupling, and $\Delta_c\sim U$, in strong coupling.
As $U$ decreases, the relative errors in extrapolating the gaps 
increase; nonetheless, the resulting fits to the scaled
Lieb-Wu gap (shown as the inset of Fig.\ \ref{liebwu}) are
still satisfied within error bars.    

In summary, within a simple model we have established that the superlattice
structure induces a shift in the critical density at which the Mott-Hubbard
transition occurs:
It increases with the ratio between free and repulsive 
layer thicknesses.
This comes about as a result of having to doubly (singly) occupy the free 
(repulsive) sites, in order to keep electronic motion to a minimum, i.e.,
to that resulting from quantum fluctuations.
These insulating states have interesting properties:
(i) fermions on free sites act as mediators of the magnetic exchange
interaction between fermions on repulsive sites (`superexchange');
(ii) if properly scaled, the charge gap seems to follow a universal curve
(reducing to that for the homogeneous system at half filling), for all
superlattices.
Experimental realizations of (effectively) one-dimensional systems 
undergoing Mott-Hubbard transitions are, by now, well established in
organic compounds\cite{expt}.
In Bechgaard salts, for instance, organic complexes are stacked along a
given direction; their large anisotropy is responsible for the one-dimensional 
character of the system. 
Within closely related families of these compounds, some are insulators 
whereas others are metals, and  
an interesting possibility would be to stack layers of these different
complexes, forming a one-dimensional superlattice, and investigate the 
evolution of the metal-insulator transitions with the superlattice structure;
we hope the scenario proposed here stimulates experimental studies along these
lines.

\acknowledgments
The authors are grateful to M.\ A.\ Continentino, J.\ d'Al\-bu\-quer\-que 
e Castro, G.\ Japaridze, and E.\ M\"uller-Hartmann for enlightening
discussions and suggestions.
Financial support from the Brazilian Agencies FINEP, CNPq and CAPES 
is also gratefully acknowledged.
The authors are also grateful to Laborat\'orio Nacional de 
Computa\c c\~ao Cient\'\i fica (LNCC) for the use of their
computational facilities.

\end{document}